# Electrochemical lithium intercalation in nanosized manganese oxides

P. Strobel[1]*, C. Darie[1], F. Thiéry[1], A. Ibarra-Palos[1†], M. Bacia[1], O. Proux[2] and J.B. Soupart[3]

[1] *Centre National de la Recherche Scientifique, Laboratorire de Cristallographie,*
*BP166, 38042 Grenoble Cedex 9, France*

[2] *Laboratoire de Géophysique Interne et Tectonophysique, UMR CNRS - Université Joseph Fourier, 1381, rue de la Piscine, 38400 Saint-Martin-d'Hères, France*

[3] *Erachem-Comilog, B-7333 Tertre, Belgium*

* *Now at Instituto de Investigaciones en Materiales, Universidad Nacional Autónoma de México, A.P. 70-360, Ciudad Universitaria, Coyoacan 04510, México, D.F.*

**Abstract**

X-ray amorphous manganese oxides were prepared by reduction of sodium permanganate by lithium iodide in aqueous medium ($MnO_x$-I) and by decomposition of manganese carbonate at moderate temperature ($MnO_x$-C). TEM showed that these materials are not amorphous, but nanostructured, with a prominent spinel substructure in $MnO_x$-C. These materials intercalate lithium with capacities up to 200 mAh/g at first cycle (potential window 1.8-4.3 V) and 175 mAh/g at 100$^{th}$ cycle. Best performances for $MnO_x$-C are obtained with cobalt doping. Potential electrochemical spectroscopy shows that the initial discharge induces a 2-phase transformation in $MnO_x$-C phases, but not in $MnO_x$-I ones. EXAFS and XANES confirm the participation of manganese in the redox process, with variations in local structure much smaller than in known long-range crystallized manganese oxides. X-ray absorption spectroscopy also shows that cobalt in $MnO_x$-C is divalent and does not participate in the electrochemical reaction.

* Corresponding Author. e-mail: strobel@grenoble.cnrs.fr. Phone: 33+ 476 887 940, Fax 33+ 476 881 038.

**Introduction**

Manganese oxides are among the most attractive positive electrode materials for lithium batteries. Because of their advantages compared to cobalt in terms of cost and non-toxicity, they have been the object of numerous studies regarding especially the $LiMn_2O_4$ (spinel-type) and $LiMnO_2$ (layered) compounds. However they present severe failures due to instability with respect to electrolyte oxidation when stored in warm conditions, which precluded their practical use up to now.

Recently, non-crystalline manganese oxides have attracted increased attention, especially since the report of Kim and Manthiram [1] of remarkable electrochemical performances for an oxi-iodide prepared by the reaction of sodium permanganate with lithium iodide in non-aqueous medium. More recently [2], Ibarra-Palos et al. showed that X-ray amorphous products could be obtained by reacting sodium permanganate with various reducing agents in aqueous medium ($Cl^-$, $I^-$, hydrogen peroxide, oxalic acid) [2]. The iodide route gave the best product in terms of ease of dehydration and electrochemical performances. Other so-called amorphous manganese oxides for lithium batteries have been reported recently. They were obtained either from permanganates with various reducing agents such as fumaric acid [3] or potassium borohydride [4], or by oxidation of aqueous $Mn^{2+}$ ion in specific conditions, especially with hypochlorite or hydrogen peroxide [5, 6] . Unlike other crystallized forms of manganese oxide, these disordered phases do not tend to convert into Li-Mn-O spinel on cycling.

A much simpler reaction, which was investigated earlier, is the decomposition of manganese carbonate. Feltz et al [7] indeed showed that, before the formation of the classical crystallized oxides such as $Mn_2O_3$, the decomposition of manganese carbonate included a step around 400°C where an unidentified, X-ray amorphous product formed.

X-ray amorphous materials present specific problems in terms of characterization of structure and reaction mechanisms, especially regarding the evidence of the topotactic, intercalation-type nature of reactions. We intend to show here on the X-ray amorphous manganese oxide the benefit of using a variety of experimental tools such as electron diffraction and X-ray absorption spectroscopy and electrochemical spectroscopy. In this paper, the contributions of these techniques will be illustrated on two kinds of disordered manganese oxides, those obtained by the permanganate-lithium iodide on one hand, by manganese carbonate decomposition on the other. These two types of materials will be called "$MnO_x$-I" and ""$MnO_x$-C" throughout this paper, respectively. Original aspects of the present study include (i) the direct inclusion of conductive carbon at the synthesis stage for $MnO_x$-I, and the study of the influence of copper or cobalt doping on the performances of $MnO_x$-C.

**Experimental**

*Synthesis procedure.* – $MnO_x$-I samples were prepared by the redox reaction between sodium permanganate and lithium iodide (Aldrich reagents). All reactions were carried out at room temperature in a 6-fold excess $Li^+$ with respect to $NaMnO_4$ concentration. Preparation details have been published elsewhere [2,8]. A new variant of this synthetic route consists in including conductive carbon black directly in the aqueous reaction medium, in order to obtain an intimate mixture oxide-carbon and to improve the electrical conductivity of the positive electrode. To this end, 250 mg of carbon black (Y50A grade, SNNA, Berre, France) was added to the LiI solution under stirring.

MnO-C samples were obtained either from commercial manganese carbonate, or from mixed manganese and M carbonates (M = Cu or Co) freshly coprecipitated from aqueous solution in the presence of lithium carbonate. The conditions of formation of X-ray amorphous oxides by thermal decomposition in air were determined, and the most favourable temperature was found to be 370°C [9]. All products were thoroughly washed with distilled water, filtered and dried under vacuum at temperatures $\geq$ 150°C before use in lithium cells.

*Chemical and structural characterization.* – Samples were studied by X-ray diffraction (XRD) using a Siemens D-5000 diffractometer with Cu K$\alpha$ radiation. The Mn, Li and Na contents (for $MnO_x$-I samples) were determined by atomic absorption analysis, and the I/Mn, Co/Mn and Cu/Mn ratios determined by EDX analysis in a JEOL 840 scanning electron microscope (SEM). The manganese oxidation state was determined by standard oxalate/permanganate volumetric titration.
Transmission electron microscope (TEM) studies were carried out using a Philips CM300 microscope operated at 300 kV (resolution 1.9 Å). Samples for TEM observation were ground under acetone and deposited on copper grids covered with thin holey carbon films.

*Electrochemical measurements.* – Electrochemical tests were carried out in liquid electrolyte using Swagelok-type batteries. Cathodic pastes were a composite mixture of active oxide, carbon black (20 wt. %) and PTFE binder (10 wt.%). The electrolyte was a commercial solution of $LiPF_6$ in EC-DMC 1:2 (Merck Co.). Negative electrodes were 200 µm-thick lithium foil (Metall Ges., Germany). Electrochemical cells were assembled in a glove box under argon with ≤ 1 ppm $H_2O$. Electrochemical studies were carried out using a MacPile Controller (Bio-Logic, Claix, France) allowing conventional galvanostatic cycling as well as slow step-potential electrochemical spectroscopy (SPES) [10], typically carried out with 10 mV per 30 mn steps.

*X-ray absorption spectroscopy.* – X-ray Absorption Spectroscopy experiments were performed at the CRG-FAME beamline (BM30B) at the European Synchrotron Radiation Facility storage ring in Grenoble, operating in 16 bunchs mode at 6 GeV [11]. Spectra were recorded in transmission mode at the Mn K edge, using a double-crystal Si(220) monochromator. The intensities of the incident and transmitted beams were measured using Si diodes. Samples for XAS measurements were pellets of diameter 10 mm made of the oxide diluted in boron nitride in appropriate proportions to give an optimum absorption jump. For the sample studied after electrochemical discharge, the thickness of the electrode pellet was adjusted to give the appropriate absorption and used as is on the beamline after opening the battery. The absolute energy scale was calibrated using a Mn metal foil ($E_K$=6.539keV) and carefully checked for each spectrum by measuring the absorption of the metallic reference, using the transmitted beam threw the samples as an incident beam for the pure Mn foil. More experimental details have been reported elsewhere [9,12].

EXAFS oscillations $\chi(k)$ were extracted from the raw data using the Athena program [13]. All the analysis were performed with the $k^2\chi(k)$ signals. The Fourier Transform (FT) of the $k^2\chi(k)$ signal was performed over the 2.6 – 11.9 Å k-range to obtain the so-called radial distribution pseudo-function, which displays peaks roughly characteristic of each shell around the central Mn atoms. Filtering of the first two peaks in the FT (containing the contribution of Mn–O and/or Mn–Mn shells) was done by inverse FT over the 0.6 – 2.9 Å R-range with a Kaiser window τ = 2.5. The fitting procedure was performed on the filtered $k^2\chi(k)$ signals with the Artemis program [13].

**Results and discussion**

*1. Physico-chemical characterization*

The permanganate-iodide route yields compounds with typical alkali contents (expressed as molar ratios with respect to manganese) 0.06-0.08 Na + 0.46-0.55 Li [2]. The resulting $MnO_x$-I samples contain small contents of iodine (2-4 %), and have a manganese oxidation state close to +4. It should be noted that the aqueous route used with a large lithium excess allows to precipitate an almost sodium-free solid phase (Na/Mn < 0.20), unlike the non-aqueous procedure initially proposed by Kim et al., which had a Na/Mn ratio of 1.51 [1]. The carbonate route produces $MnO_x$ samples with x ≈ 1.75 (Mn oxidation state close to +3.5). The doping level used is M/Mn = 0.20 (M = Co, Cu) and was confirmed by further chemical analysis. All these compouds are X-ray amorphous, as shown in Figure 1.

The specific surface areas (BET) are in the ranges 20-45 and 100-110 $m^2/g$ for the series $MnO_x$-I and $MnO_x$-C, respectively. Both have grain sizes in the 10 nm range. In addition, $MnO_x$-C samples present a peculiar agglomeration of grains in spheres at the μm-size scale [9].

*2. Transmission electron microscopy*

Investigations at the nanometer scale show that the actual crystallite size in all these materials is quite small : Figure 2 shows that samples from both I- and C-series contain grains with edges in the 10 nm range. At higher magnification, all these materials give fringes (see Figure 3) and diffraction rings - albeit broad, indicating that these materials are not amorphous, but are at least partially ordered at the nanometric scale. In the $MnO_x$-C series, discrete spots (more or less well resolved) are observed. The $MnO_x$-I SAED patterns (Figure 4a) give mostly d-spacings at 2.50 ± 0.02 and 2.05 ± 0.02 Å, which correspond to classical interreticular distances found in most structures built up from octahedral $Mn^{+4}$-O arrangements (spinel, rutile, birnessite, epsilon-$MnO_2$.etc [14]). $MnO_x$-C (Figure 4b) gave more structural information, although no clear dot-pattern from an unique diffracting grain could be obtained. The list of d-spacings recorded could be indexed best in a face-centered cubic cells with a ≈ 8.10 Å. It can therefore be concluded that $MnO_x$-C samples are nanostructured compounds based on a spinel substructure.

*3. Electrochemical intercalation properties*

The first two cycles of typical compounds of both series are shown in Figure 5. The initial capacities are > 155 mAh/g for all samples and correspond quantitatively to the reduction to $Mn^{3+}$. The charge-discharge curves are smooth S-shaped, centered around 2.8–3.0 V, in agreement with previous studies on amorphous or nanometric manganese oxides [1-3,6]. One notes, however, the difference in shape between the first discharge and subsequent cyclings in sample $MnO_x$-C (Figure 5b-c). These features can be analyzed in more detail using slow-scanning voltammetry (PITT). As shown in Figure 6, this reveals an important difference in the current decay evolution with time during potential steps in the first reduction peak. For $MnO_x$-I, the current decreases exponentially throughout the reduction peak, as expected for a single-phase, diffusion-controlled reaction [15]. For $MnO_x$-C, on the contrary, the current first decreases, then stabilizes to a constant value as a function of time. This is typical of a 2-phase reaction, where kinetics are controlled by the displacement of a phase front. This pecular behaviour is observed in all $MnO_x$-C samples investigated, irrespective of doping [9]. SPES thus unambiguously shows that $MnO_x$-C materials undergo a phase transition during the first discharge, and subsequently give stable cycling on a different phase formed in-situ.

Although it was found from SAED that $MnO_x$-C materials are based on a spinel structure at the nanometric scale, they do not at all show the double voltage plateau (3V and 4V) typical of electrochemical lithium intercalation in the "crystallized" $Li_xMn_2O_4$ spinel system. One major difference with the latter is the absence of lithium in the nanometric phases. This explains the absence of a 4V intercalation step, which is due to lithium extraction from 8a sites in the initial spinel material. In addition, whereas the 3 V plateau in ordered, crystallized $Li_xMn_2O_4$ is a 2-phase reaction with poor reversibility, this is not the case in the new nanometric phases after the first discharge. So it seems that the nanometric and disordered nature of these materials can prevent or smoothly absorb the structural distorsion appearing at high $Mn^{3+}$ concentrations due to the Jahn-Teller character of the $Mn^{3+}$ ion and the subsequent reversibility problems encountered in the long-range ordered spinel systems. The effect observed is consistent with the smoothing of discharge plateaus predicted by modelling the effect of downsizing intercalation materials [16].

Figure 7 shows the evolution of capacity with cycling in mild conditions (C/20). In spite of a higher capacity in the first 15 cycles for $MnO_x$-I, this sample exhibits a constant capacity loss, and the capacity retention on a large number of cycles is best for Cu-doped or Co-doped $MnO_x$-C. The result is a 175 mAh/g capacity at 100 cycles, showing that nanometric manganese oxides give excellent performances compared to existing "crystallized" phases used or projected in lithium batteries ($LiCoO_2$, $LiNiO_2$, $LiMn_2O_4$).

*4. X-Ray absorption analysis*

EXAFS spectra were recorded on various $MnO_x$-I and $MnO_x$-C samples, and in each case on both the pristine material and cathodic pellets extracted from a lithium battery after full discharge. Co-doped $MnO_x$-C was studied at both Mn and Co edges. Standards used were λ-$MnO_2$, $LiMn_2O_4$ and $Mn_2O_3$ for Mn, $LiCoO_2$ and $GeCo_2O_4$ for Co. Considering first the near-edge spectra (XANES), Figure 8 shows results for two typical samples. The trend is the same for both $MnO_x$-I and $MnO_x$-C : XANES clearly shows a shift of the spectra towards lower energies on discharge. As shown by the comparison with standards, these results fully agrees with (i) the Mn valences ($v_{Mn}$) determined by chemical analysis for the pristine materials (close to +4 for $MnO_x$-I, close to +3.6 for $MnO_x$-C), (ii) reduction of manganese to a final $v_{Mn}$ = +3 on discharge, as expected for a lithium intercalation reaction in these materials. Interestingly, the cobalt edge in Co-doped $MnO_x$-C does not at all exhibit such an energy shift, but is shown to be purely $Co^{2+}$ and unaffected by the electrochemical discharge [9].

Regarding EXAFS, the experimental spectra of standards λ-$MnO_2$, $LiMn_2O_4$ and $Mn_2O_3$ standards were simulated to check the validity of both the analysis procedure and the used phases and amplitudes calculated from the FEFF code [13]. The main results deduced from quantitative refinements are summarized in Table 1. The results show that the local structure around manganese is very similar in all nanometric sample under study. First-shell coordination numbers (CN) are 4.8–6, and Mn-O distances are consistent with typical distances in crystallized manganese oxides with $v_{Mn}$ in the range 3.5-4. The insertion of lithium induces only small changes in the first manganese coordination shell : the Mn-O distances increase by 0.58 % and 0.26 % in $MnO_x$-I and $MnO_x$-C samples, respectively, while CN shows a slight decrease. As shown by Debye-Weller factor values, these materials are considerably more disordered at the second shell level, and this disordered is enhanced by lithium intercalation. The second shell distances also show a larger increase with lithium intercalation (up to 1.5 %). These variations should be compared to those observed in long-range ordered manganese oxides such as the spinel $LiMn_2O_4$ (Mn valence +3.5) and $Li_2Mn_2O_4$ (Mn valence +3). In this system, the intercalation on 1 Li is accompanied by a cell volume expansion of 5.3 % and a Mn-O octahedron distortion by 15 % along the c-axis. The much smaller variations observed in the new nanometric manganese oxides are vvery favourable for reversibility of the lithium insertion-extraction reaction, and consistent with the good cyclability of these new materials.

**Conclusions**

TEM and X-ray absorption techniques prove to be essential techniques in the characterization of materials with poor X-ray diffraction features. TEM shows that the new X-ray amorphous materials presented here are made of grains with dimensions in the 10 nm range, and that they are not amorphous, but nanocrystalline. The electrochemical performances are superior to those of long-range ordered manganese oxides, probably due to the nanometric and disordered character of the new materials, which ensure a better intergrain conductivity and a hgher tolerance to changes in manganese valence. X-ray absorption spectroscopy confirms the participation of manganese (but not of cobalt) in the electrochemical redox process, and shows that the changes in local structure around Mn are indeed much smaller than those observed in "crystallized" manganese oxides.

**References**

1. J. Kim and A. Manthiram, A manganese oxyiodide cathode for rechargeable lithium batteries, Nature 390 (1997), 265-267.

2. A. Ibarra-Palos, M. Anne and P. Strobel, Electrochemical lithium intercalation in disordered manganese oxides, Solid State Ionics 138 (2001), 203-212.

3. J.J. Xu, A.J. Kinser, B.B. Owens and W.H. Smyrl, Amorphous Mn dioxide: a high capacity Li intercalation host, Eletrochem Solid State Lett., 1 (1998), 1-3.

4. C.F. Tsang, J. Kim and A. Manthiram, Synthesis of manganese oxides by reduction of $KMnO_4$ with $KBH_4$ in aqueous solutions, J. Solid State Chem. 137 (1998), 28-32.

5. J. Moon, M. Awano, H. Takai and Y.J. Fujishiro,. Synthesis of nanocrystalline manganese oxide powders: Influence of hydrogen peroxide on particle characteristics, J. Mater. Res. 14 (1999), 4594-4601.

6. J.J. Xu, G. Jain and J. Yang, Effect of copper doping on intercalation properties of amorphous manganese oxides prepared by oxidation of Mn(II) precursors, Electrochem. Solid State Lett. 5 (2005), A152-155.

7. A. Feltz, W. Ludwig and C. Felbel, Uber eine amorphe Mangan(III,IV)-oxidphase und die thermische Zersetzung von $MnCO_3$, Z. Anorg. Allg. Chem. 540 (1986), 36-44.

8. A. Ibarra-Palos, P. Strobel, C. Darie, M. Bacia and J.B. Soupart, Nanosized manganese oxides as cathode material for lithium batteries: influence of carbon mixing and grinding on cyclability, J. Power Sources (2005), in press.

9. P. Strobel, C. Darie, F. Thiéry, M. Bacia, O. Proux, A. Ibarra-Palos and J.B. Soupart, Structural characterization and electrochemical properties of new nanospherical manganese oxides for lithium batteries, J. Mater. Chem, in press (accepted).

10. A.H. Thompson, Electrochemical potential spcetroscopy: a new electrochemical measurement, J. Electrochem. Soc. 126 (1979), 608-614.

11. O. Proux, X. Biquard, E. Lahera, J.-J. Menthonnex, A. Prat, O. Ulrich, Y. Soldo, P. Trévisson, G. Kapoujvan, G. Perroux, P. Taunier, D. Grand, P. Jeantet, M. Deleglise, J.-P. Roux and J.-L. Hazemann, FAME: a new beamline for X-ray absorption investigations of diluted systems of environmental, material and biological interests, Physica Scripta 115 (2005), 970-979.

12. A. Ibarra-Palos, P. Strobel, O. Proux, J.L. Hazemann, M. Anne and M. Morcrette, In situ X-ray absorption spectroscopy study of lithium inserion in a new disordered manganese oxi-iodide, Electrochimica Acta 47 (2002), 3171-3178.

13. B. Ravel and and M. Newville, Athena and Artemis: interactive graphical data analysis using IFEFFIT, Physica Scripta 115 (2005), 1007-1017.

14. R.G. Burns and V.M. Burns, Tunnelling through $MnO_2$: covalent binding and structural linkages in tetravalent manganese oxides, ECS Proc., New Orleans Meeting, The Electrochemical Society (1984) 97-111.

15. CJ. Wen, B.A. Boukamp, R.A. Huggins and W. Weppner Electrochemical determination of thermodynamic properties of LiAl, J. Electrochem. Soc. 126 (1979), 2258-2266.

16. M.N. Obrovac and J.R. Dahn, Implications of finte-size and surface effects on nanosize intercalation materials, Phys. Rev. B 61 (2000) 6713-6719.

TABLE 1. Local structure around Mn atom as determined from EXAFS data refinement. d = interatomic distance, CN = coordination number, σ = Debye-Waller factor (in Å$^2$), R = reliability factor.

| Sample | pristine sample | | | after lithium intercalation | | |
|---|---|---|---|---|---|---|
| | d (Å) | CN | σ$^2$/R | d (Å) | CN | σ$^2$/R |
| *1st shell (Mn-O)* | | | | | | |
| MnO$_x$-I | 1.878 (11) | 5.9 (12) | 0.0038 / 0.021 | 1.889 (12) | 5.4 (11) | 0.0037 / 0.018 |
| MnO$_x$—C (Cu-doped) | 1.881 (9) | 4.8 (3) | 0..0031 / 0.012 | 1.886 (12) | 4.0 (7) | 0.0033 / 0.038 |
| MnO$_x$—C (Co-doped) | 1.885 (11) | 5.7 (8) | 0.0032 / 0.017 | 1.900 (11) | 5.2 (9) | 0.0051 / 0.018 |
| *2$^{nd}$ shell (Mn-Mn)* | | | | | | |
| MnO$_x$-I | 2.862 (13) | 3.8 (10) | 0.0056 | 2.90 (20) | 5.2 (26) | 0.013 |
| MnO$_x$-C (Cu-doped) | 2.882 (14) | 4.7 (15) | 0.013 | 2.914 (25) | (not refinable) | 0.021 |
| MnO$_x$-C (Co-doped) | 2.882 (18) | 5.7 (21) | 0.012 | 2.924 (15) | (not refinable) | 0.015 |

FIGURE CAPTIONS

Figure 1. XRD diagrams of samples $MnO_x$-C and $MnO_x$-I. The top diagram is that of standard, crystallized $MnCO_3$ recorded in same conditions.

Figure 2. Typical TEM micrographs of samples $MnO_x$-I and $MnO_x$-C.

Figure 3. TEM image of samples $MnO_x$-C at high resolution, showing diffraction fringes.

Figure 4. SAED patterns of crystallite from samples $MnO_x$-I (a) and Co-doped $MnO_x$-C (b).

Figure 5. First 2 cycles of galvanostatic charge-discharge of samples $MnO_x$-I (top) and $MnO_x$-C(middle and bottom, as marked) in lithium cells at 21°C. Discharge regime C/20.

Figure 6. Incremental current evolution during the first reduction peak (discharge) in PITT at 10 mV/30 mn for samples $MnO_x$-I (top) and $MnO_x$-C (bottom).

Fig.7. Evolution of capacity of $MnO_x$-I and Co-doped $MnO_x$-C with cycling. Conditions as in Figure 5.

Fig.8. Normalized XANES of samples $MnO_x$-I (top) and $MnO_x$-C (bottom) before and after discharge, compared to those of crystallized standards (dashed lines).

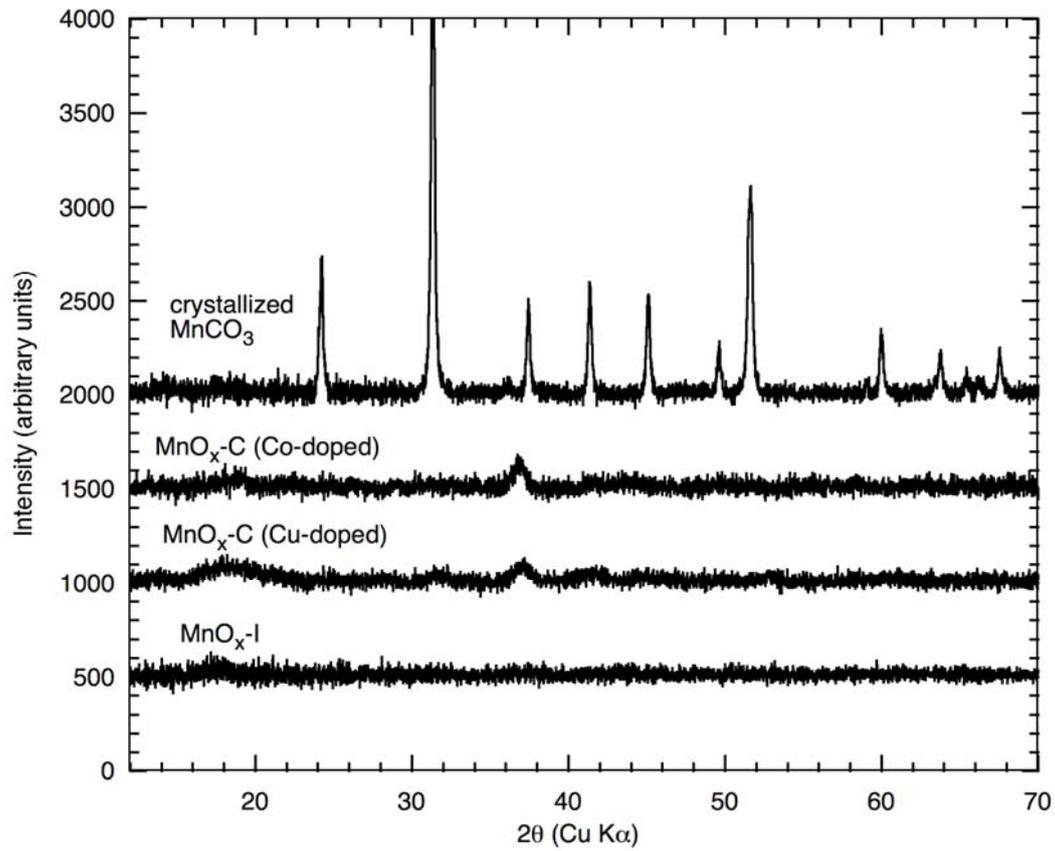

Figure 1. XRD diagrams of samples MnO$_X$-C and MnO$_X$-I. The top diagram is that of standard, crystallized MnCO$_3$ recorded in same conditions.

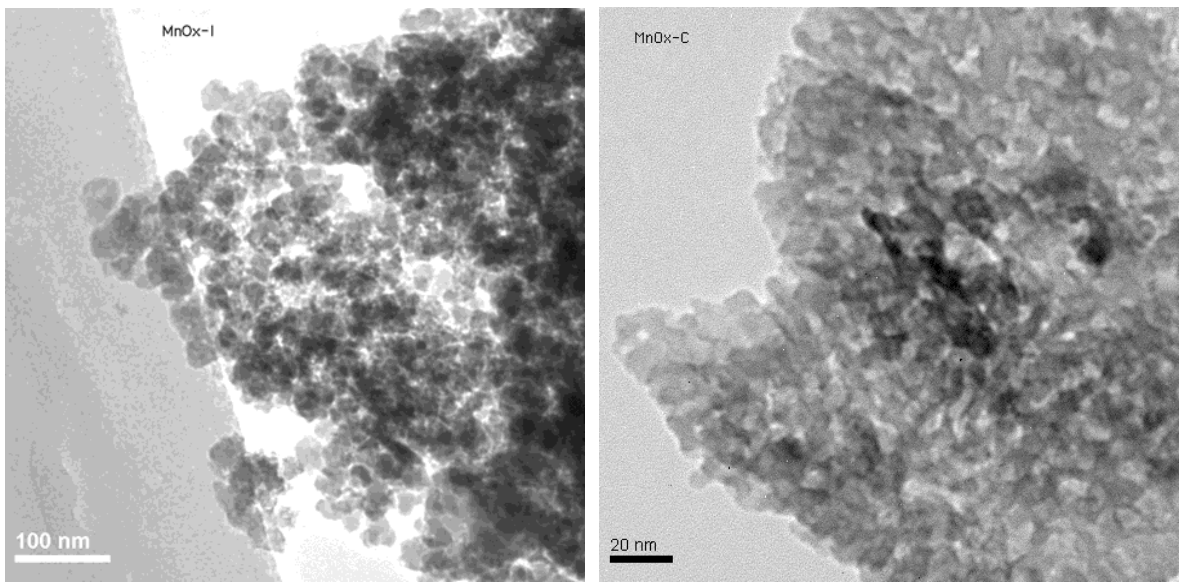

Figure 2. Typical TEM micrographs of samples MnO$_X$-I and MnO$_X$-C.

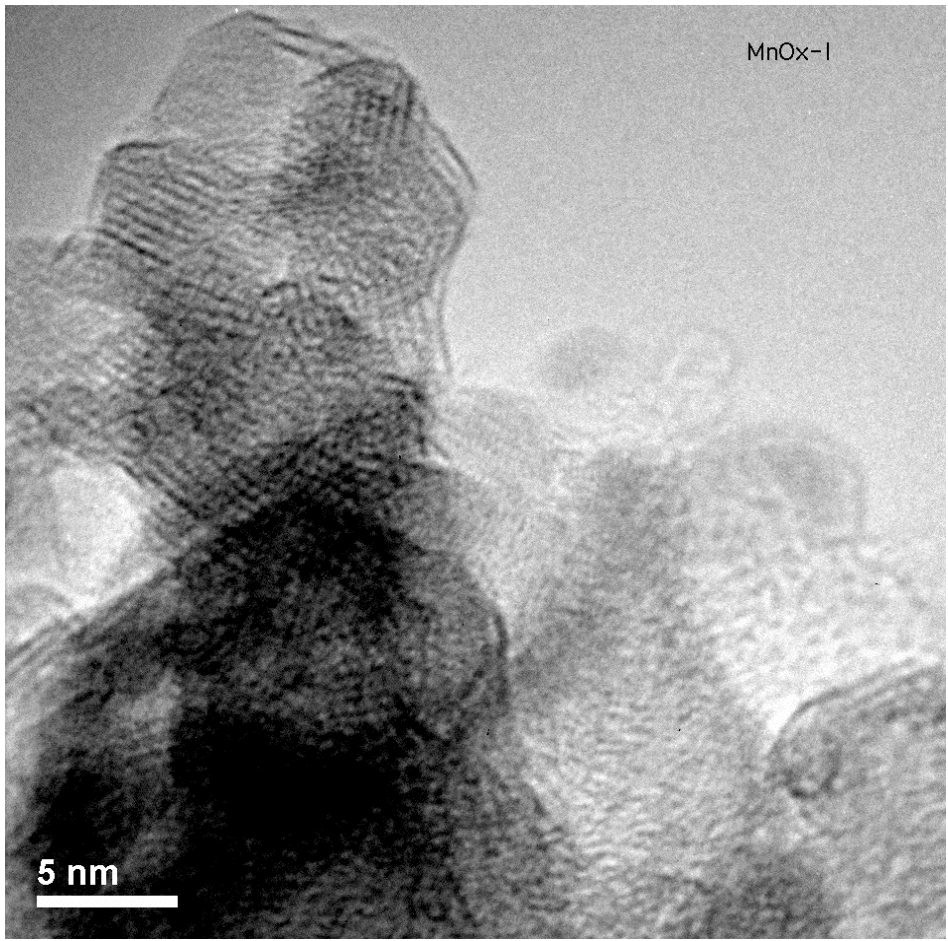

Figure 3. TEM image of samples $MnO_x$-C at high resolution, showing diffraction fringes.

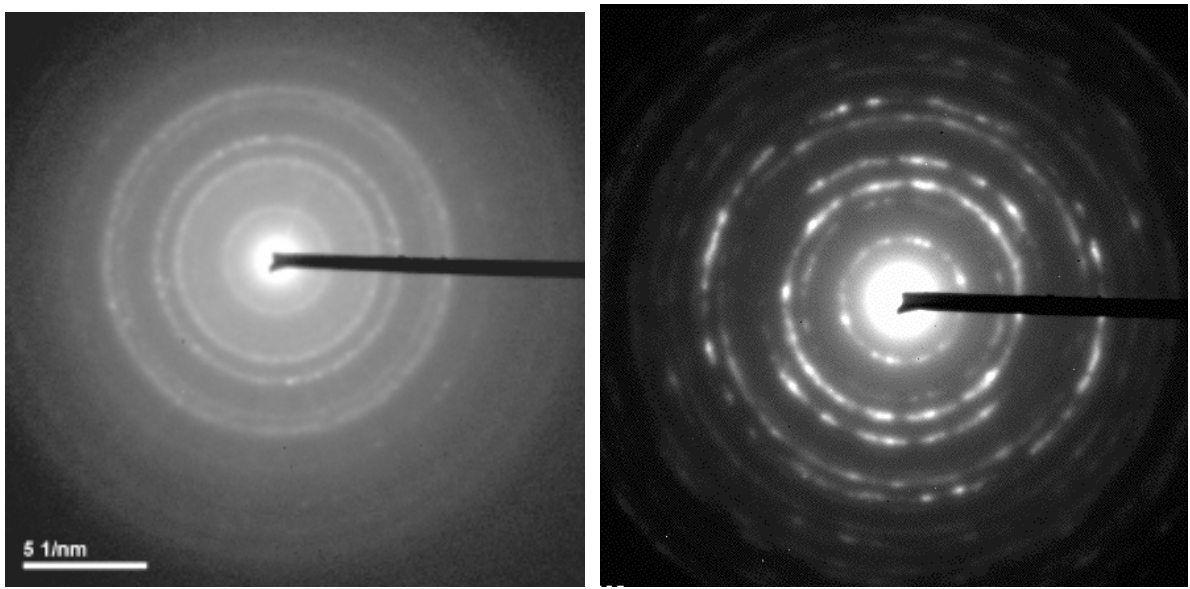

Figure 4. SAED patterns of crystallite from samples $MnO_x$-I (a) and Co-doped $MnO_x$-C (b).

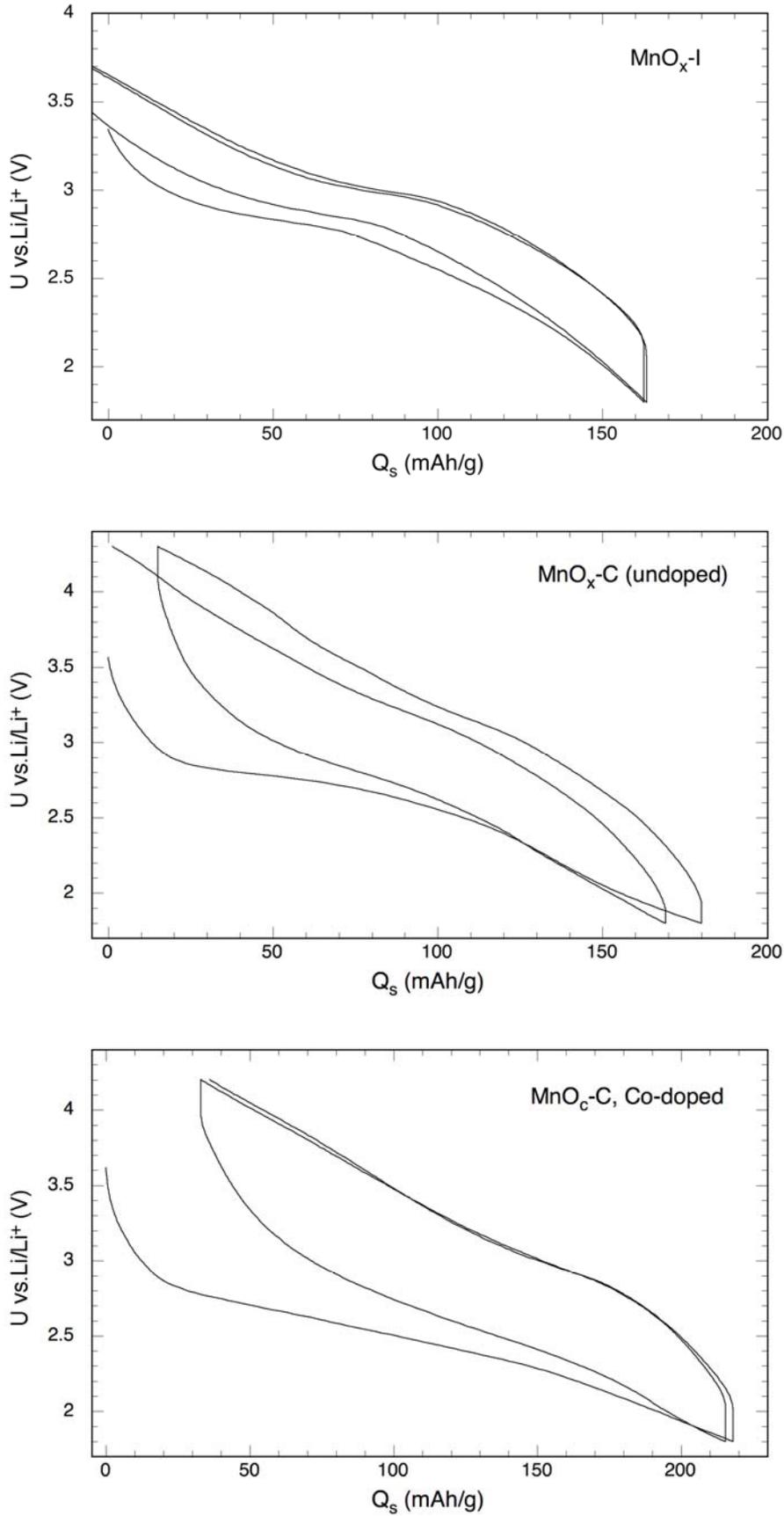

Figure 5. First 2 cycles of galvanostatic charge-discharge of samples $MnO_x$-I (top) and $MnO_x$-C (middle and bottom, as marked) in lithium cells at 21°C. Discharge regime C/20.

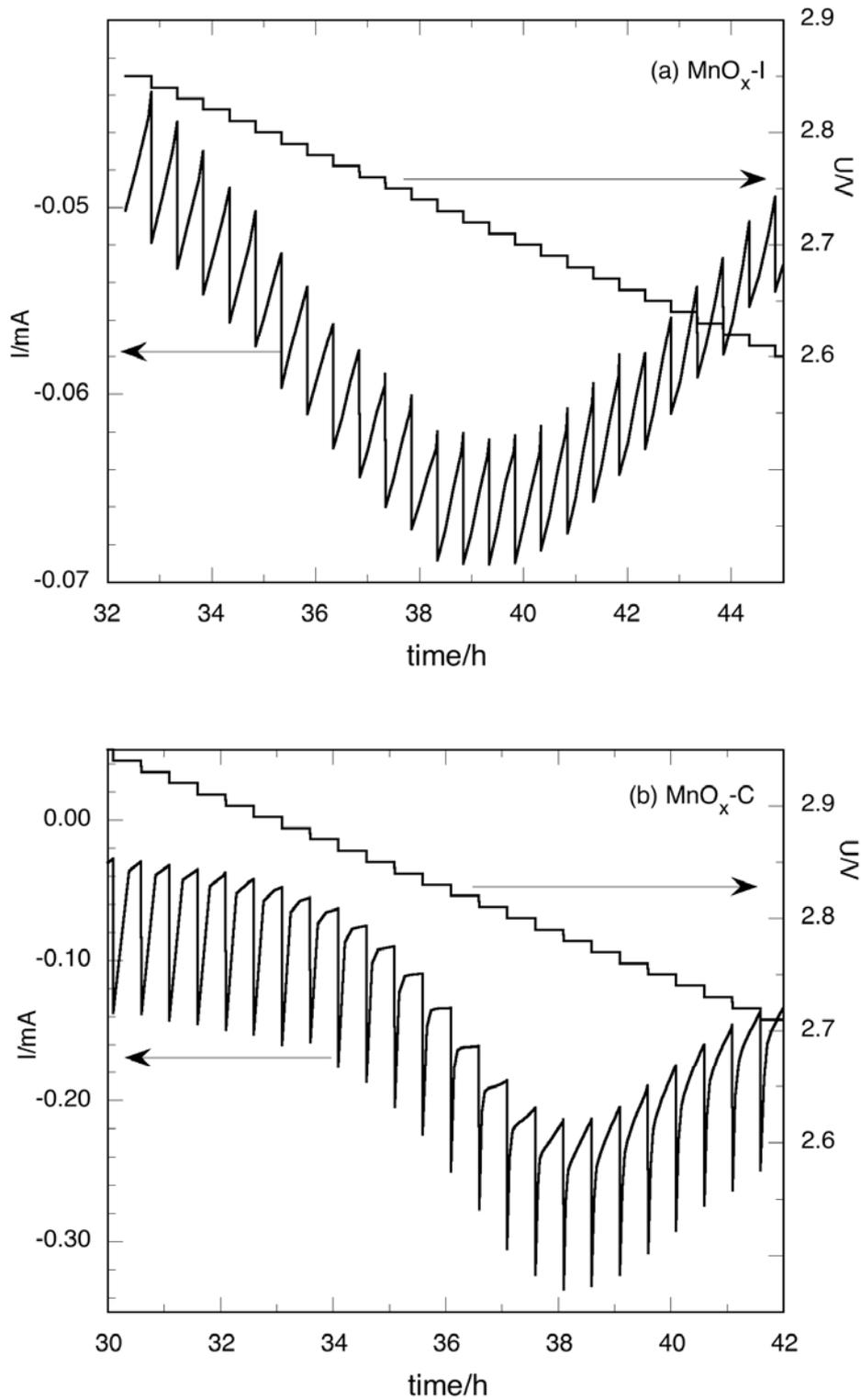

Figure 6. Incremental current evolution during the first reduction peak (discharge) in PITT at 10 mV/30 mn for samples MnO$_x$-I (top) and MnO$_x$-C (bottom).

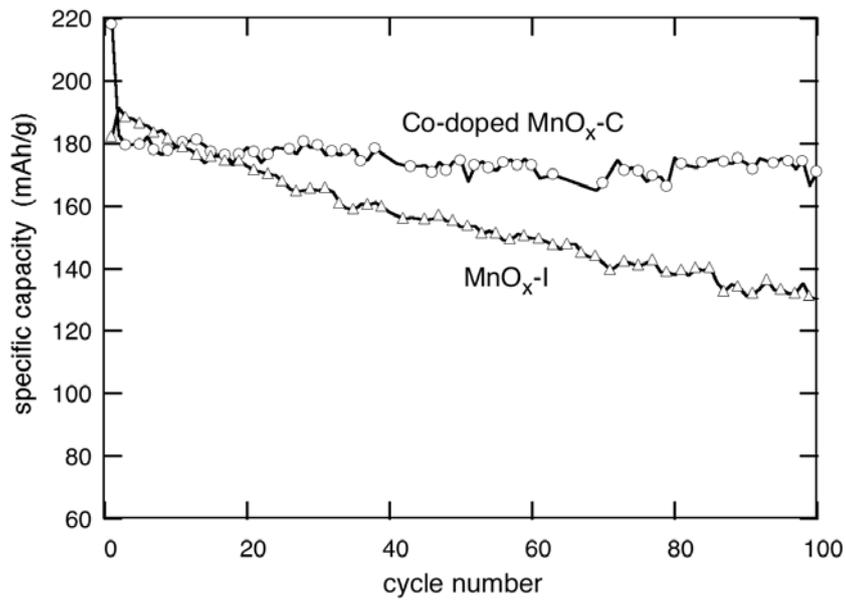

Fig.7. Evolution of capacity of MnO$_x$-I and Co-doped MnO$_x$-C with cycling. Conditions as in Figure 5.

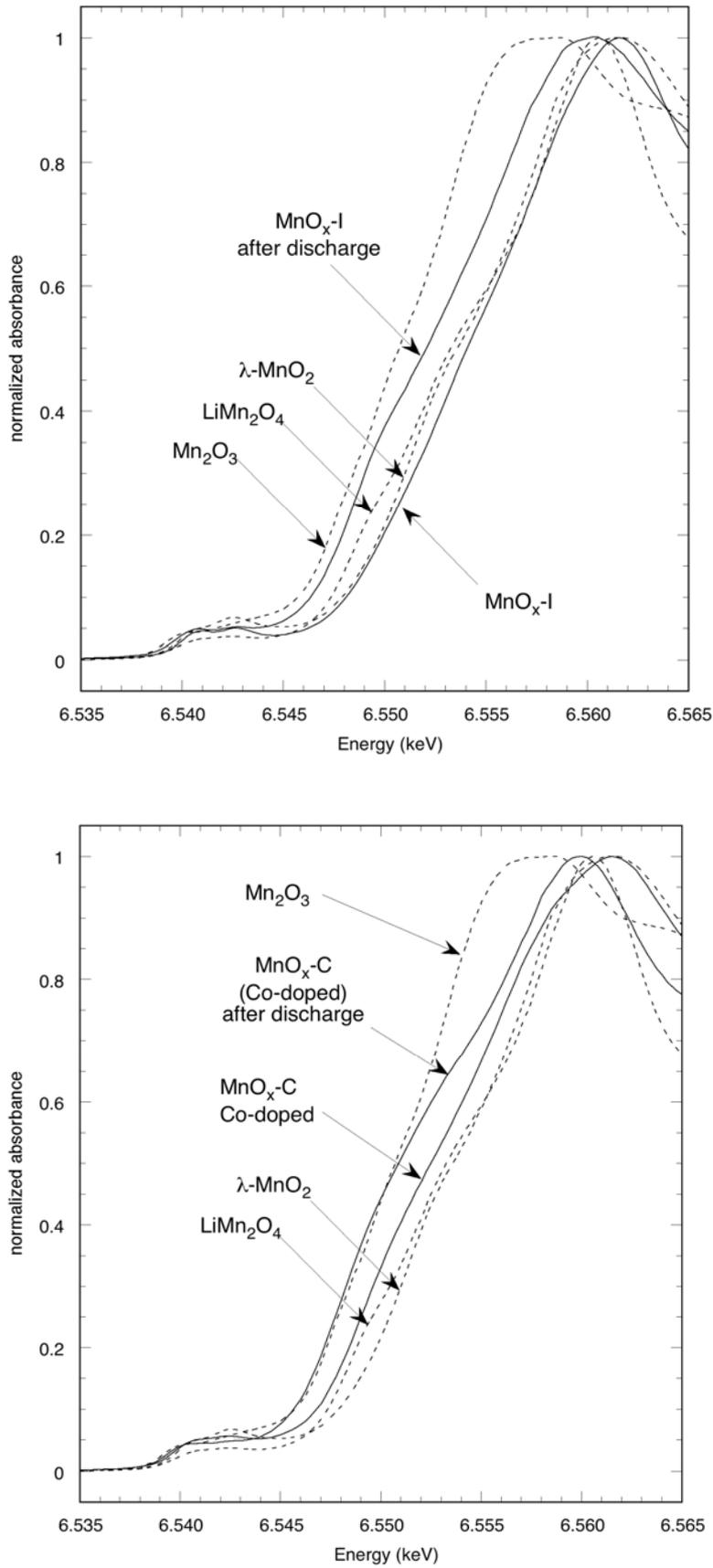

Fig.8. Normalized XANES of samples $MnO_x$-I (top) and $MnO_x$-C (bottom) before and after discharge, compared to those of crystallized standards (dashed lines).